\documentclass[english]{article}
\usepackage[T1]{fontenc}
\usepackage[latin9]{inputenc}
\usepackage{xcolor}
\usepackage{pdfcolmk}
\usepackage{array}
\usepackage{float}
\usepackage{textcomp}
\usepackage{multirow}
\usepackage{amsmath}
\usepackage{amsthm}
\usepackage{graphicx}
\usepackage{setspace}
\usepackage[numbers]{natbib}
\PassOptionsToPackage{normalem}{ulem}
\usepackage{ulem}

\makeatletter

\providecommand{\tabularnewline}{\\}
\providecolor{lyxadded}{rgb}{0,0,1}
\providecolor{lyxdeleted}{rgb}{1,0,0}

\DeclareRobustCommand{\lyxsout}[1]{\ifx\\#1\else\sout{#1}\fi}

\theoremstyle{plain}
\newtheorem{thm}{\protect\theoremname}[section]

\@ifundefined{date}{}{\date{}}
\usepackage[all]{xy}
\usepackage{morefloats}
\usepackage{placeins} 
\usepackage{lineno}
\usepackage{flushend}
\usepackage{graphicx}
\usepackage{tikz}
\usepackage{geometry}

\date{}
\newcommand{\RR}{\mathcal{R}}

\newcommand{\PP}{\mathsf{P}}

\makeatother

\usepackage{babel}
\providecommand{\theoremname}{Theorem}

\begin{document}
\title{True contextuality in a Psychophysical Experiment}
\author{Víctor H. Cervantes and Ehtibar N. Dzhafarov}
\maketitle
\begin{abstract}
Recent crowdsourcing experiments have shown that true contextuality
of the kind found in quantum mechanics can also be present in human
behavior. In these experiments simple human choices were aggregated
over large numbers of respondents, with each respondent dealing with
a single context (set of questions asked). In this paper we present
experimental evidence of contextuality in individual human behavior,
in a psychophysical experiment with repeated presentations of visual
stimuli in randomly varying conteXts (arrangements of stimuli). The
analysis is based on the Contextuality-by-Default (CbD) theory whose
relevant aspects are reviewed in the paper. CbD allows one to detect
contextuality in the presence of direct influences, i.e., when responses
to the same stimuli have different distributions in different contexts.
The experiment presented is also the first one in which contextuality
is demonstrated for responses that are not dichotomous, with five
options to choose among. CbD requires that random variables representing
such responses be dichotomized before they are subjected to contextuality
analysis. A theorem says that a system consisting of all possible
dichotomizations of responses has to be contextual if these responses
violate a certain condition, called nominal dominance. In our experiment
nominal dominance was violated in all data sets, with very high statistical
reliability established by bootstrapping.

\end{abstract}
\label{sec:behavior}Contextuality (or lack thereof) is a characteristic
of a \emph{system of random variables}. A set of random variables
forms a system if each random variable $R_{q}^{c}$ in it is uniquely
identified by its \emph{content} $q$ and its \emph{context} $c$.
The content $q$ is that which the random variable measures or responds
to, while the context $c$ is a complex of recorded conditions under
which this random variable is observed. As an example, the following
set of random variables,

\begin{equation}
\begin{array}{|c|c|c|c||c}
\hline R_{1}^{1} & R_{2}^{1} &  & R_{4}^{1} & c=1\\
\hline  & R_{2}^{2} & R_{3}^{2} &  & c=2\\
\hline R_{1}^{3} & R_{2}^{3} & R_{3}^{3} &  & c=3\\
\hline\hline q=1 & q=2 & q=3 & q=4 & \textnormal{system }\mathcal{E}
\end{array},\label{eq: E system}
\end{equation}
forms a system with three contexts and four contents.

To prevent possible misreadings, we will follow the convention adopted
in Dzhafarov and Kujala (2016a) and capitalize the distinguishing
letters in the words ``conteNt'' and ``conteXt.''

The conteNts could be, e.g., four stimuli (say, questions or light
flashes), and conteXts be defined by which two or three of them are
presented in a single trial, say, in a fixed succession. Thus, in
conteXt $c=1$, three stimuli ($q=1$, $q=2$, and $q=4$) are presented,
and each of them is being responded to in accordance with some instructions.
Depending on the arrangements, a response to a given stimulus can
be given immediately after it is presented or after all three of them
are presented \textemdash{} such experimental details are immaterial
for contextuality analysis insofar as responses and stimuli are in
a one-to-one correspondence. The responses in the conteXt $c=1$ are
the random variables $R_{1}^{1},R_{2}^{1},R_{4}^{1}$ shown in the
first row of (\ref{eq: E system}). They may be binary (e.g., Yes/No,
or I saw it/I did not see it), or they can be multi-valued ones (e.g.,
each stimulus may have a name, and the task may be to identify which
stimulus was shown). The difference between binary and more-than-binary
responses plays a central role in the present paper.

Let us explain the intuition behind the notion of contextuality using
(\ref{eq: E system}). The random variables within a given conteXt
are jointly distributed, and the marginal distribution of a given-conteNt
variable may depend on the conteXt in which it is recorded. Thus,
the distributions of $R_{2}^{2}$ and $R_{2}^{3}$ may be different,
so by knowing the distribution one can guess in which of the two conteXts,
$c=2$ or $c=3$, the conteNt $q=2$ is being responded to. This means
that the effect of a conteXt upon a distribution is \emph{information-carrying},
i.e., it is a causal influence. We call such influences \emph{direct}.
The terminology used in physics for direct influences is ``signaling,''
``disturbance,'' ``invasiveness,'' etc. (Cereceda, 2000; Leggett
\& Garg, 1985). In psychology we usually speak of ``violations of
marginal selectivity'' (Dzhafarov, 2003; Dzhafarov \& Kujala, 2016b).
If, e.g., the conteNts in (\ref{eq: E system}) are questions, and
in each conteXt they are posed in a succession, in the order of their
values ($q=1,2,3,4$), then the response $R_{2}^{2}$ to $q=2$ in
conteXt $c=2$ may very well differ in distribution from the response
$R_{2}^{3}$ to the same $q=2$ in conteXt $c=3$, because in the
later case the respondent could have been affected by the previously
asked $q=1$. The (dis)similarity of two conteNt-sharing variables,
such as $R_{2}^{2}$ and $R_{2}^{3}$, can be measured by how often
their values could coincide \emph{had they been jointly distributed}
(de facto, they are not, because they occur in mutually exclusive
conteXts). In other words, the similarity of $R_{2}^{2}$ and $R_{2}^{3}$
is measured by the maximal value of $\Pr\left[T_{2}^{2}=T_{2}^{3}\right]$
among all jointly distributed pairs $\left\{ T_{2}^{2},T_{2}^{3}\right\} $
such that $T_{2}^{2}$ is distributed as $R_{2}^{2}$, and $T_{2}^{3}$
as $R_{2}^{3}$. Any such a pair $\left\{ T_{2}^{2},T_{2}^{3}\right\} $
is called a \emph{coupling} of $R_{2}^{2}$ and $R_{2}^{3}$, and
the couplings with the maximal value of $\Pr\left[T_{2}^{2}=T_{2}^{3}\right]$
are called \emph{maximal}. We can find maximal couplings for all other
conteNt-sharing pairs $\left\{ R_{q}^{c},R_{q}^{c'}\right\} $. For
some of them we may expect no distributional differences (in our example
with questions it could be, e.g., $R_{1}^{1},R_{1}^{3}$, as in both
these cases $q=1$ is asked first), and then the maximal value of
$\Pr\left[T_{2}^{2}=T_{2}^{3}\right]$ will be $1$. This is the case
of traditional interest in quantum physics. However, generally, both
in physics and psychology, differences in distributions of conteNt-sharing
random variables should be expected and taken into account. Direct
influence is, of course, a form of conteXt-dependence, but it is very
different from what is considered \emph{contextuality} in the proper
sense of the word. The latter is detected in the system by showing
that the just mentioned maximal couplings of the conteNt-sharing pairs
are not compatible with the joint distributions of the random variables
within conteXts. In other words, a system is contextual if the joint
distributions within conteXts force the conteNt-sharing pairs across
conteNts to be more dissimilar than they could be if taken without
the conteXts. While direct influences exerted by conteNts are causal
(information-carrying), true contextuality is of a correlational,
non-causal nature.\footnote{To prevent objections, direct influences are \emph{defined} in our
theory as the differences in distributions, so one cannot speak of
``hidden'' influences (Filk, 2015, 2016). Thus, if the variables
in system $\mathcal{E}$ are binary, $+1/-1$, and $\Pr\left[R_{1}^{1}=1\right]=\Pr\left[R_{1}^{3}=1\right]=0.5$,
one can imagine that ``in reality'' conteXt $c=3$ somehow acts
upon the ``potential values'' of $R_{1}^{3}$ reversing their signs,
$R_{1}^{3}\rightarrow-R_{1}^{3}$, without changing the distribution.
However, this is not considered a ``direct influence,'' because
in the given system of random variables these unnoticeable changes
do not carry information. If one can actually observe the changes
$R_{1}^{3}\rightarrow-R_{1}^{3}$, the system of random variables
one deals with changes dramatically, and the CbD analysis then changes
accordingly (Dzhafarov, Cervantes, \& Kujala, 2017; Dzhafarov \& Kon,
2018; Dzhafarov \& Kujala, 2018).} More rigorous definitions are given below, in Section 1.

To provide historical perspective, contextuality (without using this
term at first) was introduced in quantum physics by Bell (1964, 1966)
and Kochen and Specker (1967). They demonstrated that one could meaningfully
address, using only observable measurements, the question famously
discussed in Bohr's (1935) critique of Einstein, Podolsky, and Rosen
(1935). The question is whether all measurement outcomes in a system
of measurements can be presented as being determined by some ``hidden''
random variable in a conteXt-independent way, i.e., using conteXt-independent
mappings from the values of this hidden variable into the values of
the observed measurement outcomes. With the work of Fine (1982a, b)
and Suppes and Zanotti (1981), it became clear that contextuality
can also be formulated in terms of the (non)existence of certain \emph{joint
distributions involving random variables recorded in different conteXts}.
Although some researchers disagree (Griffiths, 2017), this seems to
have become a common way of understanding contextuality (Abramsky,
Barbosa, Kishida, Lal, \& Mansfield, 2015; Abramsky, \& Brandenburger,
2011; Araújo, Quintino, Budroni, Cunha, \& Cabello, 2013; Budroni,
2016; Cabello, 2013; Khrennikov, 2008; Klyachko, Can, Binicioglu,
\& Shumovsky, 2008; Kurzynski, Ramanathan, \& Kaszlikowski, 2012;
Liang, Spekkens, \& Wiseman, 2011; Ramanathan, Soeda, Kurzynski, \&
Kaszlikowski, 2012). Probabilistic underpinnings of this understanding
have been critically examined by Khrennikov (2000a, b; 2001) and Dzhafarov
and Kujala (2016a; 2017a). Irrespective of the debated issues and
disagreements, however, contextuality analysis has been moved from
physics to probability theory, making it apparent that random variables
in contextuality analysis need not represent quantum measurements,
they can also be, e.g., responses of biological organisms to stimuli.
However, the search for contextuality in psychology was frustrated
by the fact that all behavioral systems of random variables exhibit
strong direct influences, whereas the theory of contextuality in quantum
mechanics, until recently, was only developed for \emph{consistently
connected systems}, those in which conteNt-sharing random variables
have identical distributions. When direct influences are taken into
account, a large body of experimental data collected in search of
contextuality can be shown to exhibit no contextuality (Dzhafarov,
\& Kujala, 2014; Dzhafarov, Kujala, Cervantes, Zhang, \& Jones, 2016;
Dzhafarov, Zhang, \& Kujala, 2015). Nevertheless two very recent series
of experiments unequivocally demonstrate that behavioral data (simple
conjoint choices made by people) can be represented by contextual
systems of random variables (Basieva, Cervantes, Dzhafarov, \& Khrennikov,
in press; Cervantes \& Dzhafarov, 2018). These experiments dealt with
responses aggregated over large pools of people, with each person
making choices within a single conteXt.

This paper presents the first experimental evidence of contextuality
in \emph{individual human behavior}. In the experiment presented below,
each of the three participants made repeated choices in a series of
randomized conteXts. A similar experiment, with essentially the same
stimuli and similar instructions, has been conducted before, and analyzed
in two different ways (Cervantes \& Dzhafarov, 2017a, b): both these
analyses revealed no contextuality in the data. The main difference
of that experiment from the present one is that in the former all
choices were binary, whereas in the present experiment each choice
was made among five options. This is an important difference in the
theory presented below.

\section{Contextuality-by-Default Theory}

\subsection{\label{subsec:Generalities}Generalities}

\emph{A system of random variables} is defined as a set of double-indexed
random variables
\begin{equation}
\mathcal{R}=\left\{ R_{q}^{c}:c\in C,q\in Q,q\prec c\right\} ,\label{eq: system general}
\end{equation}
where $C$ is a set of \emph{conteXts}, $Q$ is a set of \emph{conteNts},
and $q\prec c$ (or $c\succ q)$ is read ``conteNt $q$ is recorded
in conteXt $c$''.\footnote{Here and throughout, we conveniently confuse $R_{q}^{c}$ and $\left(R_{q}^{c},c,q\right)$,
so that, e.g., $\left\{ R_{q}^{c},R_{q'}^{c}\right\} $ consists of
two random variables even if $R_{q}^{c}\equiv R_{q'}^{c}$, the same
measurable function. Also, we follow the common tradition of conveniently
confusing functions $R_{q}^{c}$ with their values.} Examples of a conteNt $q$ (the ``thing'' being measured or responded
to) are particle's spin in a given direction in a Hilbert space, or
a question asked of a person. Examples of a conteXt $c$ may be subsets
of conteNts measured ``together'' (simultaneously or sequentially),
or different conditions associated with a given subset of conteNts
(e.g., the order in which two fixed questions are asked). The corresponding
$R_{q}^{c}$ would then be the spin value (say, ``up'' or ``down'')
along axis $q$ in a given set $c$ of measured properties, or the
response (say, ``yes'' or ``no'') to question $q$ asked before
or after another question, $q'$, with $c=\left(q',q\right)$. As
a \emph{random variable}, $R_{q}^{c}$ is a measurable function from
a probability space $\left(X^{c},\varXi^{c},\pi^{c}\right)$ to a
measurable space $\left(Y_{q},\varUpsilon_{q}\right)$, with the usual
meaning of the components. The probability space $\left(Y_{q},\varUpsilon_{q},p_{q}^{c}\right)$
induced by this function is the \emph{distribution} of $R_{q}^{c}$.
The indices show that $\left(X^{c},\varXi^{c},\pi^{c}\right)$ is
common to all $R_{q}^{c}$ within a conteXt $c$, i.e., all such $R_{q}^{c}$
are jointly distributed, reflecting the fact that their realizations
are empirically linked. Put differently, for any $c\in C$, the set
\begin{equation}
R^{c}=\left\{ R_{q}^{c}:q\in Q,q\prec c\right\} 
\end{equation}
can be viewed as a random variable. It is a principle of CbD that
any $R_{q}^{c},R_{q'}^{c'}$ with $c\not=c'$ are \emph{stochastically
unrelated}, i.e., $(X^{c},\varXi^{c},\pi^{c})\not=(X^{c'},\varXi^{c'},\pi^{c'})$,
reflecting the fact that conteXts are mutually exclusive, so no pairing
of the values of $R_{q}^{c}$ and $R_{q'}^{c'}$ is defined. In particular,
the variables in
\begin{equation}
\mathcal{R}_{q}=\left\{ R_{q}^{c}:c\in C,q\prec c\right\} \label{eq: connection}
\end{equation}
for a given $q$ are not jointly distributed. However, the distributions
of any $R_{q}^{c},R_{q}^{c'}$ in $\mathcal{R}_{q}$ always share
the same measurable space, $\left(Y_{q},\varUpsilon_{q}\right)$,
reflecting the fact that $R_{q}^{c}$ and $R_{q}^{c'}$ have the same
conteNt (i.e., they measure or respond to the same ``thing'').

The next definition is a modification of the usual one (Thorisson,
2000), to better suit our purposes. A (probabilistic) \emph{coupling}
of an indexed set of random variables $\left\{ V_{i}\right\} _{i\in I}$
is an identically indexed set of jointly distributed random variables
$\left\{ W_{i}\right\} _{i\in I}$ such that, for any subset $I'\subseteq I$,
if the elements of $\left\{ V_{i}\right\} _{i\in I'}$ are jointly
distributed, then $\left\{ W_{i}\right\} _{i\in I'}\overset{dist}{=}\left\{ V_{i}\right\} _{i\in I'}$
(the same distribution). In particular, a coupling of a system $\RR$
in (\ref{eq: system general}) is a set 
\begin{equation}
S=\left\{ S_{q}^{c}:c\in C,q\in Q,q\prec c\right\} 
\end{equation}
of jointly distributed random variables, such that, for all $c\in C$,
\begin{equation}
R^{c}=\left\{ R_{q}^{c}:q\in Q,q\prec c\right\} \overset{dist}{=}\left\{ S_{q}^{c}:q\in Q,q\prec c\right\} =S^{c}.
\end{equation}
Returning to our example (\ref{eq: E system}), the following matrix
of jointly distributed random variables (or simply, the following
random variable) $E$,
\begin{equation}
\begin{array}{|c|c|c|c||c}
\hline S_{1}^{1} & S_{2}^{1} &  & S_{4}^{1} & c=1\\
\hline  & S_{2}^{2} & S_{3}^{2} &  & c=2\\
\hline S_{1}^{3} & S_{2}^{3} & S_{3}^{3} &  & c=3\\
\hline\hline q=1 & q=2 & q=3 & q=4 & \textnormal{coupling }E
\end{array},\label{eq: E system-1}
\end{equation}
is a coupling of $\mathcal{E}$ if $S^{c}\overset{dist}{=}R^{c}$
for $c=1,2,3$.

Let $\PP_{\max}$ be the following statement, well-defined (in the
sense of being true or false) for any two jointly distributed random
variables $A,B$:
\begin{equation}
\PP_{\max}\left(A,B\right)=\textnormal{\textquotedblleft\ensuremath{\Pr\left[A=B\right]} is maximal possible, given the distributions of \ensuremath{A} and \ensuremath{B.}\textquotedblright}\label{eq: =00005CPPmax}
\end{equation}
If a coupling $\left\{ S_{q}^{c},S_{q}^{c'}\right\} $ of two conteNt-sharing
random variables $R_{q}^{c}$ and $R_{q}^{c'}$ satisfies this statement,
it is called a maximal coupling of $R_{q}^{c}$ and $R_{q}^{c'}$.
The system $\RR$ is \emph{noncontextual} if $\RR$ has a coupling
$S$ in which any $\left\{ S_{q}^{c},S_{q}^{c'}\right\} $ is a \emph{maximal
coupling} of $R_{q}^{c}$ and $R_{q}^{c'}$. Otherwise, if such a
coupling $S$ does not exist, the system is \emph{contextual}. Using
our example in (\ref{eq: E system-1}), system $\mathcal{E}$ is noncontextual
if and only if among all its couplings $E$ one can find at least
one in which all equalities $S_{1}^{1}=S_{1}^{3}$, $S_{2}^{1}=S_{2}^{2}$,
$S_{2}^{2}=S_{2}^{3}$, and $S_{3}^{2}=S_{3}^{3}$ occur with the
maximal probability allowed by their individual distributions. Thus,
if $R_{1}^{1}$ and $R_{1}^{3}$ are dichotomous, $+1/-1$, with $\Pr\left[R_{1}^{1}=1\right]=p$
and $\Pr\left[R_{1}^{3}=1\right]=q$, then the maximal possible probability
of $S_{1}^{1}=S_{1}^{3}$ is $1-\left|p-q\right|$. Obviously, any
subsystem of a noncontextual system (obtained by deleting some of
the random variables) is noncontextual, or, equivalently, any system
with a contextual subsystem is contextual.

\subsection{\label{subsec:Dichotomous-random-variables}Dichotomous random variables}

Most systems of traditional interest consist of \emph{dichotomous}
random variables. Among basic properties of such systems one should
mention the following (Dzhafarov, 2017; Dzhafarov, Cervantes, \& Kujala,
2017; Dzhafarov \& Kujala, 2017a, b).
\begin{description}
\item [{(P1)}] Adding to or removing from a system a deterministic random
variable (attaining a single value with probability 1), or a variable
that does not share its conteXt or its conteNt with other variables,
does not change the system's (non)contextuality (in fact, does not
change the degree of contextuality, but we do not discuss this notion
here).
\item [{(P2)}] A set of conteNt-sharing random variables $\mathcal{R}_{q}=\left\{ R_{q}^{c}:c\in C,q\prec c\right\} $
always has a \emph{unique} coupling such that any two of its elements
satisfy $\PP_{\max}$. (Such a coupling is referred to as a \emph{multimaximal
coupling}).
\item [{(P3)}] $T_{q}=\left\{ T_{q}^{c}:c\in C,q\prec c\right\} $ is a
multimaximal coupling of $\RR_{q}$ if and only if, for any $\left\{ c_{1},\ldots,c_{k}\right\} \subseteq C$,
the probability of $T_{q}^{c_{1}}=\ldots=T_{q}^{c_{k}}$ is maximal
among all couplings of $\left\{ R_{q}^{c_{1}},\ldots,R_{q}^{c_{k}}\right\} $.
\item [{(P4)}] If $\mathcal{R}_{q}=\left\{ R_{q}^{c_{1}},\ldots,R_{q}^{c_{l}}\right\} $
is enumerated so that $\Pr\left[R_{q}^{c_{1}}=1\right]\leq\ldots\leq\Pr\left[R_{q}^{c_{l}}=1\right],$
then $T_{q}$ is a multimaximal coupling of $\RR_{q}$ if and only
if $\Pr\left[T_{q}^{c_{i}}=T_{q}^{c_{i+1}}\right]$ is maximal for
$i=1,\ldots,l-1$ among all possible couplings of $\RR_{q}$.
\end{description}
Especially important in quantum-mechanical applications are cyclic
systems of ranks $n=2,3,\ldots$. Denoting by $\oplus1$ cyclic clockwise
shift $1\mapsto2,\ldots,n-1\mapsto n,n\mapsto1$ (and by $\ominus1$
the opposite shift), a cyclic system of rank $n$ has conteXts $c=1,\ldots,n$,
conteNts $q=1,\ldots,n$, and consists of dichotomous ($+1/-1$) random
variables $\left\{ R_{i}^{i},R_{i\oplus1}^{i}:i=1,\ldots,n\right\} $.
Some examples of such systems are: for $n=2$, question order effects
(Wang \& Busemeyer, 2013; Wang, Solloway, Shiffrin, \& Busemeyer,
2014); for $n=3$, the Suppes-Zanotti (Suppes \& Zanotti, 1981), original
Bell (1964), and Leggett-Garg (Leggett \& Garg, 1985) systems in quantum
mechanics, and simple decision making systems in cognition (Asano,
Hashimoto, Khrennikov, Ohya, \& Tanaka, 2014; Basieva et al., in press);
for $n=4$, the EPR/Bohm-Bell-CHSH systems (Bell, 1966; Bohm \& Aharonov,
1957; Clauser \& Horne, 1974; Clauser, Horne, Shimony, \& Holt, 1969;
Fine, 1982a, b), and decision making and psychophysical systems (Bruza,
Kitto, Nelson, \& McEvoy, 2009; Bruza, Kitto, Ramm, \& Sitbon, 2015;
Cervantes \& Dzhafarov, 2017, 2018); for $n=5$, the KCBS system (Klyachko
et al., 2008; Lapkiewicz, Li, Schaeff, Langford, Ramelow, Wie\'{s}niak,
\& Zeilinger, 2011); for $n>5$, some psychophysical systems (Zhang
\& Dzhafarov, 2016). The main theoretical result here is\\
\begin{thm}[Kujala \& Dzhafarov, 2016]
 \emph{A cyclic system of rank $n$ is contextual if and only if
(denoting expected value by $\left\langle \cdot\right\rangle $)}
\begin{equation}
\max_{\iota_{1},\ldots,\iota_{k}\in\left\{ -1,1\right\} ,\prod_{i=1}^{n}\iota_{i}=-1}\sum_{i=1}^{n}\iota_{i}\left\langle R_{i}^{i}R_{i\oplus1}^{i}\right\rangle -\left(n-2\right)-\sum_{i=1}^{n}\left|\left\langle R_{i}^{i}\right\rangle -\left\langle R_{i}^{i\ominus1}\right\rangle \right|>0.\label{eq: criterion}
\end{equation}
\end{thm}

[\noindent] Prior to Kujala and Dzhafarov (2016), this general result
was conjectured and proved for small values of $n$ (Dzhafarov, Kujala,
\& Larsson, 2015; Kujala \& Dzhafarov, 2015). The special case of
this result for consistently connected systems had been proved, by
very different means, in (Araújo et al., 2013).

We do not have analogous closed-form criteria for non-cyclic systems,
but the theory here is well-developed. There is a general linear programming
method for establishing contextuality or lack thereof in any given
system with finite sets $C$ and $Q$ and dichotomous random variables
(Dzhafarov \& Kujala, 2016a; Dzhafarov, Cervantes, \& Kujala, 2017)
(in fact, the method would work for any categorical random variables,
but the CbD approach does not require this, see Section \ref{subsec:Arbitrary-random-variables}).
The problem is reduced to a certain underdetermined system of linear
equations, 
\begin{equation}
\mathbf{M}\mathbf{Q}=\mathbf{P}.\label{eq:MQ=00003DP}
\end{equation}
Here, $\mathbf{P}=\left(1,\overset{\#1}{\ldots},\overset{\#2}{\ldots}\right)$,
where \#1 denotes all probabilities characterizing the distributions
within the conteXts (e.g., $\Pr\left[R_{1}^{1}=1,R_{2}^{1}=1,R_{3}^{1}=-1\right]$),
and \#2 denotes all probabilities characterizing the maximal couplings
$\left\{ T_{q}^{c},T_{q}^{c'}\right\} $ of the separate conteNt-sharing
pairs (e.g., $\Pr\left[T_{2}^{1}=1,T_{2}^{2}=1\right]$); $\mathbf{Q}$
is a vector of probabilities (summing to 1) for all possible values
of the hypothetical coupling $S$; and $\mathbf{M}$ is a Boolean
matrix with 1's in each row corresponding to values of $S$ comprising
the events whose probabilities are given in $\mathbf{P}$. The system
is noncontextual if and only if these linear equations have a solution
for $\mathbf{Q}$ with nonnegative components. The linear programming
representation of CbD naturally leads to its geometric representations
by polytopes and graph-theoretic renderings. A detailed version of
the latter was recently proposed by Amaral, Duarte, and Oliveira (2018).

\subsection{\label{subsec:Arbitrary-random-variables}Arbitrary random variables}

The current version of CbD (Dzhafarov, Cervantes, \& Kujala, 2017;
Dzhafarov \& Kujala, 2017a, b) posits that all random variables in
a system should be dichotomized before they are submitted to contextuality
analysis. One reason for this is that the property \textbf{P2} in
the previous section does not hold for non-dichotomous variables:
a multimaximal coupling need not exist, and when it does, need not
be unique. The other reason is that one expects a noncontextual systems
to remain noncontextual if some values of a random variable are ``lumped
together'' (e.g., if in $\left\{ 1,2,3,4\right\} $ one ceases to
distinguish $1$ and $2$) (Dzhafarov, Cervantes, \& Kujala, 2017).
Dichotomizations are easy if in the initial description of an empirical
domain all random variables are categorical (i.e., have unordered
finite sets of values). One then is interested in all possible dichotomizations:
an $n$-valued random variable is replaced with $2^{n-1}-1$ distinct
dichotomizations (with unordered pairs of values). For instance, if
an initial $R$ has values $\left\{ 1,2,3,4\right\} $, in contextual
analysis it is replaced with 7 jointly distributed 
\begin{equation}
\begin{array}{cc|c|c|c|c|c|c|}
 & R_{\left(1\right)} & R_{\left(2\right)} & R_{\left(3\right)} & R_{\left(4\right)} & R_{\left(5\right)} & R_{\left(6\right)} & R_{\left(7\right)}\\
\textnormal{values:} & 1\,\Vert\,2,3,4 & 2\,\Vert\,1,3,4 & 3\,\Vert\,1,2,4 & 4\,\Vert\,1,2,3 & 1,2\,\Vert\,3,4 & 1,3\,\Vert\,2,4 & 1,4\,\Vert\,2,3
\end{array}
\end{equation}
Assume, e.g., that in system $\mathcal{E}$ of (\ref{eq: E system})
the variables for $q=1$ have 4 values, variables for $q=3$ have
3 values, and the other two variables are binary. Dichotomization
of the system then transforms it into
\begin{equation}
\begin{array}{|c|c|c|c|c|c|c|c|c|c|c|c||c}
\hline R_{1(1)}^{1} & R_{1(2)}^{1} & R_{1(3)}^{1} & R_{1(4)}^{1} & R_{1(5)}^{1} & R_{1(6)}^{1} & R_{1(7)}^{1} & R_{2}^{1} &  &  &  & R_{4}^{1} & c=1\\
\hline  &  &  &  &  &  &  & R_{2}^{2} & R_{3(1)}^{2} & R_{3(2)}^{2} & R_{3(3)}^{2} &  & c=2\\
\hline R_{1(1)}^{3} & R_{1(2)}^{3} & R_{1(3)}^{3} & R_{1(4)}^{3} & R_{1(5)}^{3} & R_{1(6)}^{3} & R_{1(7)}^{3} & R_{2}^{3} & R_{3(1)}^{3} & R_{3(2)}^{3} & R_{3(3)}^{3} &  & c=3\\
\hline\hline q=1(1) & 1(2) & 1(3) & 1(4) & 1(5) & 1(6) & 1(7) & 2 & 3(1) & 3(2) & 3(3) & 4 & \textnormal{system }\mathcal{E}^{*}
\end{array}
\end{equation}
where the numbers in parentheses encode different dichotomizations.
The procedure effectively splits old conteNts into new conteNts. The
size of the system increases only in visual appearance, because in
each row of $\mathcal{E}^{*}$ the support of the joint distribution
is precisely the same as in system $\mathcal{E}$. The original system
is considered contextual if its dichotomization is contextual. The
main result here is\\

\noindent\textbf{Theorem} (Dzhafarov, Cervantes, \& Kujala, 2017).
\emph{A system of categorical random variables (before dichotomization)
is contextual if, for some $\left(q,c,c'\right)$, neither of $R_{q}^{c},R_{q}^{c'}$
nominally dominates the other. }\\

It is this theorem that we use to analyze the experiment below. The
meaning of nominal dominance is as follows: given $A$ and $B$ with
the same set of values $\left\{ 1,\ldots,k\right\} $, $A$ \emph{nominally
dominates} $B$ if the inequality $\Pr\left[A=i\right]<\Pr\left[B=i\right]$
holds for no more than one value of $i=1,\ldots,k$ (i.e., if $\Pr\left[A=i\right]\geq\Pr\left[B=i\right]$
for at least $k-1$ of them). Thus, among the pairs of probability
distributions below,
\begin{equation}
\begin{array}{ccc}
\begin{array}{r|c|c|c|c|c|}
\textnormal{(i)}\qquad\textnormal{Values} & \textnormal{1} & \textnormal{2} & \textnormal{3} & \textnormal{4} & \textnormal{5}\\
\hline \textnormal{probabilities for \ensuremath{A}:} & 0.1 & 0.2 & 0.2 & 0.5 & 0\\
\textnormal{probabilities for \ensuremath{B}:} & 0.1 & 0.2 & 0.2 & 0.5 & 0
\end{array}, &  & \begin{array}{r|c|c|c|c|c|}
\textnormal{(ii)}\qquad & \textnormal{1} & \textnormal{2} & \textnormal{3} & \textnormal{4} & \textnormal{5}\\
\hline \textnormal{probabilities for \ensuremath{A}:} & 0.1 & 0.2 & 0.2 & 0.5 & 0\\
\textnormal{probabilities for \ensuremath{B}:} & 0.2 & 0.1 & 0.2 & 0.5 & 0
\end{array},\\
\\
\begin{array}{r|c|c|c|c|c|}
\textnormal{(iii)}\qquad\textnormal{Values} & \textnormal{1} & \textnormal{2} & \textnormal{3} & \textnormal{4} & \textnormal{5}\\
\hline \textnormal{probabilities for \ensuremath{A}:} & 0.1 & 0.2 & 0.2 & 0.5 & 0\\
\textnormal{probabilities for \ensuremath{B}:} & 0.5 & 0 & 0.1 & 0.4 & 0
\end{array}, &  & \begin{array}{r|c|c|c|c|c|}
\textnormal{(iv)}\qquad\textnormal{Values} & \textnormal{1} & \textnormal{2} & \textnormal{3} & \textnormal{4} & \textnormal{5}\\
\hline \textnormal{probabilities for \ensuremath{A}:} & 0.1 & 0.2 & 0.2 & 0.5 & 0\\
\textnormal{probabilities for \ensuremath{B}:} & 0.3 & 0.3 & 0.1 & 0.2 & 0.1
\end{array},
\end{array}
\end{equation}
in (i) and (ii) $A$ and $B$ nominally dominate each other, in (iii)
$A$ nominally dominates $B$, and in (iv) neither of the two random
variables nominally dominates the other.

The theorem above tells us that if we are interested in all possible
dichotomizations, we may not need to actually create them to determine
that the system is contextual. It suffices instead to find at least
one instance when neither of two original (as observed, before dichotomization)
conteNt-sharing random variables nominally dominates the other, as
in (iv) above. The condition is only sufficient but not necessary
for contextuality: if nominal dominance is found in all pairs of conteNt-sharing
random variables, the system may or may not be contextual.

\section{Double-Identification Experiment}

\subsection{Method}

\subsubsection{Participants}

\label{sec:part}Three volunteers, graduate students at Purdue University,
one female and two males (including the first author of this paper),
with normal or corrected to normal vision, participated in this study.
The experimental program was regulated by Purdue University's IRB
protocol $\#1202011876$. The participants are identified as P1, P2,
and P3 in the text below.

\subsubsection{Equipment}

\label{sec:apparatus} 
\begin{figure}
\begin{centering}
\includegraphics[scale=0.25]{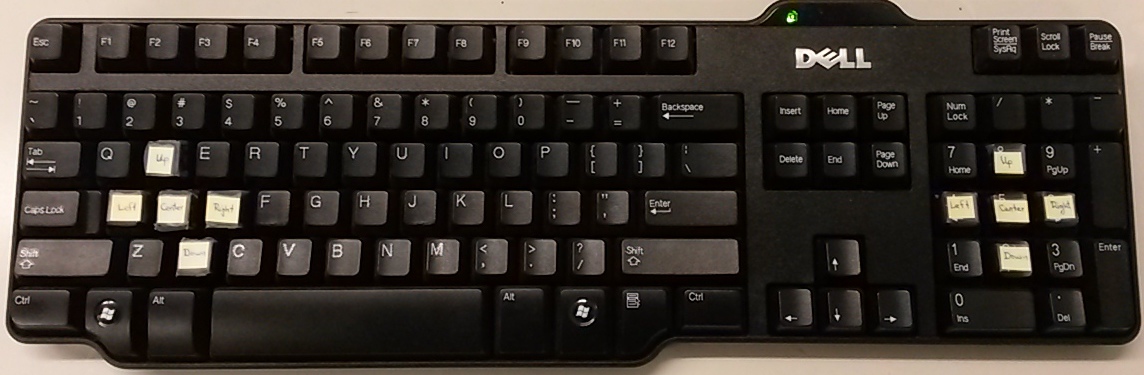}
\par\end{centering}
\caption{Layout of the keyboard with the response keys stickers for left and
right stimuli.}
\label{fig:keyboard}
\end{figure}

A personal computer was used with an Intel{\scriptsize{}®}
Core{\scriptsize{}\texttrademark{}} processor running Windows XP,
and with a $24$-in.\ monitor with a resolution of $1920\times1200$
pixels (px). The participant's head was steadied in a chin-rest with
forehead support at 90 cm distance from the monitor; at this distance
a pixel on the screen subtended $62$ sec arc. The response keys on
a US $104$-key keyboard were indicated by stickers with the corresponding
response labels (see Figure \ref{fig:keyboard}).

\subsubsection{Stimuli}

\label{sec:stim} 
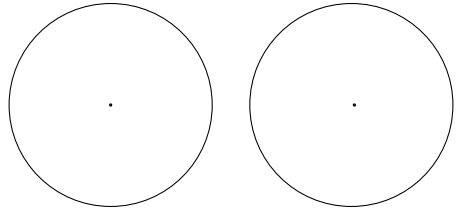
\begin{figure}[tbh]
\centering{}\begin{center}
\begin{tikzpicture}
	\draw[line width=.4] (8,6) circle[radius=1.35];
	\draw[line width=.4] (11.2,6) circle[radius=1.35];
	\filldraw[thick] (8,6) circle[radius=.2pt];
	\filldraw[thick] (11.24,6) circle[radius=.2pt];
\end{tikzpicture}
\par\end{center} \caption{An example of the stimuli in experiment (in reversed contrast and
not to scale). In the left circle the dot is in the center, in the
right one it is shifted to the right by 4 px ($\simeq4.1$min arc).
The participant's task was to identify the location of the dot in
each of the two circles by pressing corresponding keys on a keyboard.}
\label{fig:stimulus}
\end{figure}

The stimuli presented on the computer screen consisted
of two brightly grey colored circles (RGB 100-100-100) on a black
background, with their centers 320 px apart horizontally, each circle
having the radius of $135$ px and 4 px wide circumference. Each circle
contained within it a dot of 4 px in diameter, that could be located
in the circle's center or 4 px away from it, in the left, right, upward
or downward direction. An example of the stimuli is shown in Figure~\ref{fig:stimulus}.

\subsubsection{Procedure}

\label{sec:proc} In each trial the participant was asked to indicate,
for each circle, whether the dot was in its center or shifted in one
of the four directions (up, down, left, or right). The responses were
given by pressing in any order and holding together two designated
keys, one for each location in each circle, as shown in Fig. \ref{fig:stimulus}.
The stimuli were displayed until both keys were pressed. Then, the
dots in each circle disappeared, and the next pair of dots appeared
$600$ ms later. The circles, with or without the dots, remained displayed
continuously throughout the experiment. (Response times were recorded
but not used in the data analysis.)

Each participant completed between $20$ and $23$ experimental sessions,
each lasting $30$ minutes and consisting of about $380$ trials recorded
and used for subsequent analysis. The experimental sessions were preceded
by two training sessions, excluded from the analysis. The first 75
trials of each training session were practice trials in which the
participants received feedback as to whether their response for each
of the two circles was correct or not. No feedback was given in the
experimental trials.

\subsection{Experimental conteXts and conteNts}

\label{sec:expCond} In each of two circles the dot presented could
be in one of 5 locations: at the center, or shifted to the left, right,
up, or down. These locations formed conteNts of the random variables
in the probabilistic description of the experiment, denoted as shown
in Table \ref{tab:allocation}. The same table shows that the $5\times5$
pairs of locations of the two dots formed $25$ conteXts. In each
experimental session, all conteXts were presented {[}close-to-{]}equal
numbers of times (about 15).

\begin{table}
\caption{Notation used for the conteXts and the conteNts: $c$, $l$, $r$,
$u$, and $d$ denote that the dot is, respectively, in the center,
shifted to the left, to the right, up, or down. The 25 conteXts are
denoted $cc$, $cu$, $du$, etc., the left (right) symbol indicating
the location of the dot in the left (respectively, right) circle.
To denote conteNts, the location of a dot is shown on the left (for
the left circle) or on the right (for the right circle) of a dash:
thus, $c\textnormal{-}$ denotes the dot in the center of the left
circle, $\textnormal{-}l$ denotes the dot shifted to the left in
the right circle, etc.\label{tab:allocation}}

\centering{}{\small{}}%
\begin{tabular}{|r|r|c|c|c|c|c|}
\cline{3-7} \cline{4-7} \cline{5-7} \cline{6-7} \cline{7-7} 
\multicolumn{1}{r}{} &  & \multicolumn{5}{c|}{{\small{}Right circle conteNts}}\tabularnewline
\cline{3-7} \cline{4-7} \cline{5-7} \cline{6-7} \cline{7-7} 
\multicolumn{1}{r}{} &  & {\small{}$\left(\textnormal{-}c\right)$} & {\small{}$\left(\textnormal{-}l\right)$} & {\small{}$\left(\textnormal{-}r\right)$} & {\small{}$\left(\textnormal{-}u\right)$} & {\small{}$\left(\textnormal{-}d\right)$}\tabularnewline
\hline 
\multirow{5}{*}{{\small{}Left circle conteNts}} & {\small{}Center $\left(c\textnormal{-}\right)$} & {\small{}$cc$} & {\small{}$cl$} & {\small{}$cr$} & {\small{}$cu$} & {\small{}$cd$}\tabularnewline
\cline{2-7} \cline{3-7} \cline{4-7} \cline{5-7} \cline{6-7} \cline{7-7} 
 & {\small{}Left $\left(l\textnormal{-}\right)$} & {\small{}$lc$} & {\small{}$ll$} & {\small{}$lr$} & {\small{}$lu$} & {\small{}$ld$}\tabularnewline
\cline{2-7} \cline{3-7} \cline{4-7} \cline{5-7} \cline{6-7} \cline{7-7} 
 & {\small{}Right $\left(r\textnormal{-}\right)$} & {\small{}$rc$} & {\small{}$rl$} & {\small{}$rr$} & {\small{}$ru$} & {\small{}$rd$}\tabularnewline
\cline{2-7} \cline{3-7} \cline{4-7} \cline{5-7} \cline{6-7} \cline{7-7} 
 & {\small{}Up $\left(u\textnormal{-}\right)$} & {\small{}$uc$} & {\small{}$ul$} & {\small{}$ur$} & {\small{}$uu$} & {\small{}$ud$}\tabularnewline
\cline{2-7} \cline{3-7} \cline{4-7} \cline{5-7} \cline{6-7} \cline{7-7} 
 & {\small{}Down $\left(d\textnormal{-}\right)$} & {\small{}$dc$} & {\small{}$dl$} & {\small{}$dr$} & {\small{}$du$} & {\small{}$dd$}\tabularnewline
\hline 
\end{tabular}{\small\par}
\end{table}

For each session, each trial was randomly assigned to one of the conditions
in Figure~\ref{tab:allocation}. The number of experimental sessions
was chosen so that the expected number of experimental trials in each
of the conteXts was at least $300$. This number of observations was
chosen based on Cepeda Cuervo, Aguilar, Cervantes, Corrales, Díaz,
and Rodríguez (2008), whose results show that coverage errors with
respect to nominal values are below $1\%$ for most confidence intervals
for proportions with $n>300$.

The system of random variables describing the experiment is shown
in Figure \ref{fig:positionSystem}.

\begin{figure}[H]
\begin{centering}
{\scriptsize{}}%
\begin{tabular}{c|c|c|c|c|c|c|c|c|c|c|}
\multicolumn{1}{c}{} & \multicolumn{1}{c}{$c\textnormal{-}$} & \multicolumn{1}{c}{$\textnormal{-}c$} & \multicolumn{1}{c}{$l\textnormal{-}$} & \multicolumn{1}{c}{$\textnormal{-}l$} & \multicolumn{1}{c}{$r\textnormal{-}$} & \multicolumn{1}{c}{$\textnormal{-}r$} & \multicolumn{1}{c}{$u\textnormal{-}$} & \multicolumn{1}{c}{$\textnormal{-}u$} & \multicolumn{1}{c}{$d\textnormal{-}$} & \multicolumn{1}{c}{$\textnormal{-}d$}\tabularnewline
\cline{2-11} \cline{3-11} \cline{4-11} \cline{5-11} \cline{6-11} \cline{7-11} \cline{8-11} \cline{9-11} \cline{10-11} \cline{11-11} 
$cc$ & $\star$ & $\star$ & $ $ & $ $ & $ $ & $ $ & $ $ & $ $ & $ $ & $ $\tabularnewline
\cline{2-11} \cline{3-11} \cline{4-11} \cline{5-11} \cline{6-11} \cline{7-11} \cline{8-11} \cline{9-11} \cline{10-11} \cline{11-11} 
$cl$ & $\star$ & $ $ & $ $ & $\star$ & $ $ & $ $ & $ $ & $ $ & $ $ & $ $\tabularnewline
\cline{2-11} \cline{3-11} \cline{4-11} \cline{5-11} \cline{6-11} \cline{7-11} \cline{8-11} \cline{9-11} \cline{10-11} \cline{11-11} 
$cr$ & $\star$ & $ $ & $ $ & $ $ & $ $ & $\star$ & $ $ & $ $ & $ $ & $ $\tabularnewline
\cline{2-11} \cline{3-11} \cline{4-11} \cline{5-11} \cline{6-11} \cline{7-11} \cline{8-11} \cline{9-11} \cline{10-11} \cline{11-11} 
$cu$ & $\star$ & $ $ & $ $ & $ $ & $ $ & $ $ & $ $ & $\star$ & $ $ & $ $\tabularnewline
\cline{2-11} \cline{3-11} \cline{4-11} \cline{5-11} \cline{6-11} \cline{7-11} \cline{8-11} \cline{9-11} \cline{10-11} \cline{11-11} 
$cd$ & $\star$ & $ $ & $ $ & $ $ & $ $ & $ $ & $ $ & $ $ & $ $ & $\star$\tabularnewline
\cline{2-11} \cline{3-11} \cline{4-11} \cline{5-11} \cline{6-11} \cline{7-11} \cline{8-11} \cline{9-11} \cline{10-11} \cline{11-11} 
$lc$ & $ $ & $\star$ & $\star$ & $ $ & $ $ & $ $ & $ $ & $ $ & $ $ & $ $\tabularnewline
\cline{2-11} \cline{3-11} \cline{4-11} \cline{5-11} \cline{6-11} \cline{7-11} \cline{8-11} \cline{9-11} \cline{10-11} \cline{11-11} 
$ll$ & $ $ & $ $ & $\star$ & $\star$ & $ $ & $ $ & $ $ & $ $ & $ $ & $ $\tabularnewline
\cline{2-11} \cline{3-11} \cline{4-11} \cline{5-11} \cline{6-11} \cline{7-11} \cline{8-11} \cline{9-11} \cline{10-11} \cline{11-11} 
$lr$ & $ $ & $ $ & $\star$ & $ $ & $ $ & $\star$ & $ $ & $ $ & $ $ & $ $\tabularnewline
\cline{2-11} \cline{3-11} \cline{4-11} \cline{5-11} \cline{6-11} \cline{7-11} \cline{8-11} \cline{9-11} \cline{10-11} \cline{11-11} 
$lu$ & $ $ & $ $ & $\star$ & $ $ & $ $ & $ $ & $ $ & $\star$ & $ $ & $ $\tabularnewline
\cline{2-11} \cline{3-11} \cline{4-11} \cline{5-11} \cline{6-11} \cline{7-11} \cline{8-11} \cline{9-11} \cline{10-11} \cline{11-11} 
$ld$ & $ $ & $ $ & $\star$ & $ $ & $ $ & $ $ & $ $ & $ $ & $ $ & $\star$\tabularnewline
\cline{2-11} \cline{3-11} \cline{4-11} \cline{5-11} \cline{6-11} \cline{7-11} \cline{8-11} \cline{9-11} \cline{10-11} \cline{11-11} 
$rc$ & $ $ & $\star$ & $ $ & $ $ & $\star$ & $ $ & $ $ & $ $ & $ $ & $ $\tabularnewline
\cline{2-11} \cline{3-11} \cline{4-11} \cline{5-11} \cline{6-11} \cline{7-11} \cline{8-11} \cline{9-11} \cline{10-11} \cline{11-11} 
$rl$ & $ $ & $ $ & $ $ & $\star$ & $\star$ & $ $ & $ $ & $ $ & $ $ & $ $\tabularnewline
\cline{2-11} \cline{3-11} \cline{4-11} \cline{5-11} \cline{6-11} \cline{7-11} \cline{8-11} \cline{9-11} \cline{10-11} \cline{11-11} 
$rr$ & $ $ & $ $ & $ $ & $ $ & $\star$ & $\star$ & $ $ & $ $ & $ $ & $ $\tabularnewline
\cline{2-11} \cline{3-11} \cline{4-11} \cline{5-11} \cline{6-11} \cline{7-11} \cline{8-11} \cline{9-11} \cline{10-11} \cline{11-11} 
$ru$ & $ $ & $ $ & $ $ & $ $ & $\star$ & $ $ & $ $ & $\star$ & $ $ & $ $\tabularnewline
\cline{2-11} \cline{3-11} \cline{4-11} \cline{5-11} \cline{6-11} \cline{7-11} \cline{8-11} \cline{9-11} \cline{10-11} \cline{11-11} 
$rd$ & $ $ & $ $ & $ $ & $ $ & $\star$ & $ $ & $ $ & $ $ & $ $ & $\star$\tabularnewline
\cline{2-11} \cline{3-11} \cline{4-11} \cline{5-11} \cline{6-11} \cline{7-11} \cline{8-11} \cline{9-11} \cline{10-11} \cline{11-11} 
$uc$ & $ $ & $\star$ & $ $ & $ $ & $ $ & $ $ & $\star$ & $ $ & $ $ & $ $\tabularnewline
\cline{2-11} \cline{3-11} \cline{4-11} \cline{5-11} \cline{6-11} \cline{7-11} \cline{8-11} \cline{9-11} \cline{10-11} \cline{11-11} 
$ul$ & $ $ & $ $ & $ $ & $\star$ & $ $ & $ $ & $\star$ & $ $ & $ $ & $ $\tabularnewline
\cline{2-11} \cline{3-11} \cline{4-11} \cline{5-11} \cline{6-11} \cline{7-11} \cline{8-11} \cline{9-11} \cline{10-11} \cline{11-11} 
$ur$ & $ $ & $ $ & $ $ & $ $ & $ $ & $\star$ & $\star$ & $ $ & $ $ & $ $\tabularnewline
\cline{2-11} \cline{3-11} \cline{4-11} \cline{5-11} \cline{6-11} \cline{7-11} \cline{8-11} \cline{9-11} \cline{10-11} \cline{11-11} 
$uu$ & $ $ & $ $ & $ $ & $ $ & $ $ & $ $ & $\star$ & $\star$ & $ $ & $ $\tabularnewline
\cline{2-11} \cline{3-11} \cline{4-11} \cline{5-11} \cline{6-11} \cline{7-11} \cline{8-11} \cline{9-11} \cline{10-11} \cline{11-11} 
$ud$ & $ $ & $ $ & $ $ & $ $ & $ $ & $ $ & $\star$ & $ $ & $ $ & $\star$\tabularnewline
\cline{2-11} \cline{3-11} \cline{4-11} \cline{5-11} \cline{6-11} \cline{7-11} \cline{8-11} \cline{9-11} \cline{10-11} \cline{11-11} 
$dc$ & $ $ & $\star$ & $ $ & $ $ & $ $ & $ $ & $ $ & $ $ & $\star$ & $ $\tabularnewline
\cline{2-11} \cline{3-11} \cline{4-11} \cline{5-11} \cline{6-11} \cline{7-11} \cline{8-11} \cline{9-11} \cline{10-11} \cline{11-11} 
$dl$ & $ $ & $ $ & $ $ & $\star$ & $ $ & $ $ & $ $ & $ $ & $\star$ & $ $\tabularnewline
\cline{2-11} \cline{3-11} \cline{4-11} \cline{5-11} \cline{6-11} \cline{7-11} \cline{8-11} \cline{9-11} \cline{10-11} \cline{11-11} 
$dr$ & $ $ & $ $ & $ $ & $ $ & $ $ & $\star$ & $ $ & $ $ & $\star$ & $ $\tabularnewline
\cline{2-11} \cline{3-11} \cline{4-11} \cline{5-11} \cline{6-11} \cline{7-11} \cline{8-11} \cline{9-11} \cline{10-11} \cline{11-11} 
$du$ & $ $ & $ $ & $ $ & $ $ & $ $ & $ $ & $ $ & $\star$ & $\star$ & $ $\tabularnewline
\cline{2-11} \cline{3-11} \cline{4-11} \cline{5-11} \cline{6-11} \cline{7-11} \cline{8-11} \cline{9-11} \cline{10-11} \cline{11-11} 
$dd$ & $ $ & $ $ & $ $ & $ $ & $ $ & $ $ & $ $ & $ $ & $\star$ & $\star$\tabularnewline
\cline{2-11} \cline{3-11} \cline{4-11} \cline{5-11} \cline{6-11} \cline{7-11} \cline{8-11} \cline{9-11} \cline{10-11} \cline{11-11} 
\end{tabular}{\scriptsize\par}
\par\end{centering}
\caption{The conteNt-conteXt system of measurements for the double detection
experiment. The cell corresponding to conteXt $xy$ and conteNt $z$
(with $z$ being $x\textnormal{-}$ or $\textnormal{-}y$), if it
contains a star, represents the random variable $R_{z}^{xy}$; the
absence of a star means that conteNt $z$ was not measured in conteXt
$xy$. For instance, $xy=cc$ and $z=c\textnormal{-}$ define a random
variable $R_{c\textnormal{-}}^{cc}$. There are two random jointly
distributed variables, $R_{x\textnormal{-}}^{xy}$ and $R_{\textnormal{-}y}^{xy}$,
in each conteXt $xy$, and their joint distribution is defined by
the probabilities: $\Pr\left[R_{x\textnormal{-}}^{xy}=j,R_{\textnormal{-}y}^{xy}=k\right]$
where $j,k\in$\{center, left, right, up, down\}. \label{fig:positionSystem}}
\end{figure}

\subsection{Results}

\label{sec:results}

The complete set of results obtained in the experiment (excluding
training sessions) is stored in \textquotedbl Contextuality in a
psychophysical double-identification experiment\textquotedbl , https:\slash\slash doi.org\slash 10.7910\slash DVN\slash FCT9VO.
The data used in the analysis of the nominal dominance condition are
shown in Tables A1, A2, and A3, placed in Appendix. These tables show
the estimated probabilities with which each of the three participants
responded in each of five possible ways (center, left, right, up,
and down) to the left stimulus and to the right stimulus, in each
of the 25 conteXts. For all participants, the nominal dominance condition
fails for at least one pair of random variables for each of the conteNts.
This means that, for all three participants, the pattern of the results
indicates contextuality.

To assess the reliability of these results, we generated $100000$
bootstrap resamples for each participant: each bootstrap resample
was generated by independently selecting, with replacement, a random
sample from (and of the same size as) the responses given in the experiment
to each of the two circles in each conteXt. The proportions of resamples
in which nominal dominance was observed are presented in Table \ref{fig:bootstrap},
for each conteNt separately, and (in the bottom row of the table)
for all conteNts simultaneously. Note that it is the latter that matters
for our analysis: the system may be noncontextual only if nominal
dominance is satisfied for all pairs of conteNt-sharing random variables.
This was observed for none of the resamples and none of the participants.
We can model this situation, for each participant, as a sequence of
100000 binomial trials with zero successes. If $p$ denotes the probability
of this happening (let us label this as a ``success''), We can model
the results, for each participant, as a Bernoulli sequence of length
100000, with probability of a ``success'' (overall compliance with
nominal dominance) being $p$, and the observed number of successes
being zero. The exact $99.999\%$ Clopper-Pearson (Clopper \& Pearson,
1934) confidence interval for $p$ is $[0,0.00012]$. We can clearly
dismiss the possibility that our data result from random perturbations
of a pattern that satisfies nominal dominance.

\begin{table}
\caption{Bootstrap estimates of the probabilities for the systems to satisfy
the nominal dominance condition.\label{fig:bootstrap}}
\smallskip{}

\centering{}$\begin{tabular}{lrrr}
\hline  ConteNt  &  P1  &  P2  &  P3 \\
\hline  \ensuremath{c}-  &  0.038  &  0.000  &  0.000 \\
 \ensuremath{l}-  &  0.000  &  0.000  &  0.224\\
 \ensuremath{r}-  &  0.000  &  0.000  &  0.003\\
 \ensuremath{u}-  &  0.429  &  0.000  &  0.023\\
 \ensuremath{d}-  &  0.002  &  0.000  &  0.001\\
 -\ensuremath{c}  &  0.412  &  0.000  &  0.000 \\
 -\ensuremath{l}  &  0.019  &  0.000  &  0.385\\
 -\ensuremath{r}  &  0.000  &  0.000  &  0.015\\
 -\ensuremath{u}  &  0.566  &  0.001  &  0.034\\
 -\ensuremath{d}  &  0.001  &  0.000  &  0.000\\
\hline  \ensuremath{\begin{array}{c}
 \textnormal{Overall,}\\
 \textnormal{for all contents} 
\end{array}}  &  0.000  &  0.000  &  0.000 \\
\hline  
\end{tabular}$
\end{table}

\section{Discussion}

\label{sec:discussion}

Based on the CbD analysis of many published experiments in none of
which contextuality was found, it was tempting to hypothesize that
all behavioral systems were noncontextual (Dzhafarov, Kujala, Cervantes,
Zhang, \& Jones, 2016; Dzhafarov, Zhang, \& Kujala, 2015; Zhang \&
Dzhafarov, 2016). This hypothesis was rejected by recent crowdsourcing
experiments (Basieva et al., in press; Cervantes \& Dzhafarov, 2018),
but the question remained open as to whether contextuality can also
be observed in individual human behavior. In the crowdsourcing experiments
the stimuli were questions to be answered in one of two ways. In such
an experiment a repeated presentation of a question to the same person
cannot be viewed as a repeated recording of the same random variable,
because the person would most likely remember her previous answers
and repeat them not to contradict herself, or would deliberately vary
them due to the phenomenon of satiation. Therefore, to investigate
contextuality in a within-subject paradigm, one has to use stimuli
that do not have any distinguishing characteristics by which they
can be remembered. Thus, if a variety of weak flashes varying in intensity
are judged in terms of ``I have seen it'' or ``I have not seen
it,'' there is no way the observer may remember seeing a particular
flash before, unless this flash was seen with probability 1. Analogously,
in our experiment, there was no way a participant could remember seeing
a specific dot position in one of the circles, as no position was
identified perfectly.

A previously conducted experiment (Cervantes \& Dzhafarov, 2017a,
b), similar to the one presented in this paper, revealed no contextuality,
i.e., all conteXt-dependence in it could be attributed to direct influences.
In that experiment the dots within two circles could vary on three
levels (center, up, down) and the responses were dichotomous: ``in
the center'' or ``not in the center.'' As it turns out, switching
to questions with five possible answers (and increasing the number
of conteNts to five to match them) changed the system from noncontextual
to contextual.

The overall conteXt-dependence in our experiment means that a given
location $q$ of the dot in a circle is judged differently for different
locations $q'$ of the dot in the other circle. This direct influence
of $q'$ on responses to $q$ manifests itself in the changing distribution
of the responses to $q$ as $q'$ changes. The contextuality of the
system, however, shows that these direct influences cannot account
for the entire situation: the changes in the identity of the random
variable representing the responses to $q$ in different conteXts
are greater than warranted by their distributional differences. This
is another way of stating the definition of a contextual system, according
to which the joint distributions of the random variables within conteXts
force conteNt-sharing random variables (responses to the same $q$
at different $q'$) to be more dissimilar than warranted by the difference
in their distributions.

The relationship between the two forms of conteXt-dependence in a
contextual system, direct influences and contextuality proper, is
a complex issue of which we have very little knowledge at present.
A remarkable fact is that this relationship seems to be different
in systems of binary random variables (at least in cyclic systems,
mentioned in Section \ref{subsec:Dichotomous-random-variables}) and
in systems of multivalued random variables. As is evident from (\ref{eq: criterion}),
the direct influences and contextuality in a cyclic system are antagonistic.
Direct influences in (\ref{eq: criterion}) are represented by 
\begin{equation}
\sum_{i=1}^{n}\left|\left\langle R_{i}^{i}\right\rangle -\left\langle R_{i}^{i\ominus1}\right\rangle \right|,
\end{equation}
and as this quantity increases, the value of the left-hand-side expression
in (\ref{eq: criterion}) decreases, making the system less likely
to be contextual. In our present experiment the situation is more
complex. Direct influences here are responsible for the differences
between the distributions 
\[
\begin{array}{r|c|c|c|c|c|}
\textnormal{responses to \ensuremath{q} in context \ensuremath{qq'}:} & \textnormal{center} & \textnormal{left} & \textnormal{right} & \textnormal{up} & \textnormal{down}\\
\hline \textnormal{probabilities:} & p_{1} & p_{2} & p_{3} & p_{4} & p_{5}
\end{array}
\]
and
\[
\begin{array}{r|c|c|c|c|c|}
\textnormal{responses to \ensuremath{q} in context \ensuremath{qq''}:} & \textnormal{center} & \textnormal{left} & \textnormal{right} & \textnormal{up} & \textnormal{down}\\
\hline \textnormal{probabilities:} & p'_{1} & p'_{2} & p'_{3} & p'_{4} & p'_{5}
\end{array}.
\]
In the absence of all direct influences, i.e., with $p_{i}=p'_{i}$
for all $i$, the nominal dominance is trivially satisfied. This does
not mean that the system in noncontextual, but its contextuality will
have to be established by other means, generally, by solving the linear
programming task (\ref{eq:MQ=00003DP}). Direct influences must be
present to break the nominal dominance relation and thereby allow
us to establish contextuality ``easily.'' More work is needed to
understand this relationship better.

\paragraph*{Acknowledgments.}

This research has been partially supported by AFOSR grant FA9550-14-1-0318.

\section*{References}

\begin{onehalfspace}
Abramsky, S., Barbosa, R. S., Kishida, K., Lal, R., \& Mansfield,
S. (2015). Contextuality, cohomology and paradox. \emph{Computer Science
Logic, 2015, }211-228.

Abramsky, S. \&Brandenburger, A. (2011). The sheaf\textminus theoretic
structure of non\textminus locality and contextuality. \emph{New Journal
of Physics, 13}, 113036-113075.

Amaral, B., Duarte, C., \& Oliveira, R. I. (2018). Necessary conditions
for extended noncontextuality in general sets of random variables.
\emph{Journal of Mathematical Physics, 59}, 072202.

Araújo, M., Quintino, M. T. , Budroni, C. , Cunha, M. T., \& Cabello,
A. (2013). All noncontextuality inequalities for the n-cycle scenario.
\emph{Physical Review A, 88,} 022118.

Asano, M., Hashimoto, T., Khrennikov, A., Ohya, M., \& Tanaka, T.
(2014). Violation of contextual generalization of the Leggett-Garg
inequality for recognition of ambiguous figures. \emph{Physica Scripta,
T163}, 014006.

Basieva, I., Cervantes, V. H., Dzhafarov, E. N., \& Khrennikov, A.
(in press). True contextuality beats direct influences in human decision
making. \emph{Journal of Experimental Psychology: General} (available
as arXiv:1807.05684).

Bell, J. (1964). On the Einstein-Podolsky-Rosen paradox. \emph{Physics,
1,} 195-200.

Bell, J. (1966). On the problem of hidden variables in quantum mechanics.
\emph{Review of Modern Physics, 38}, 447-453.

Bohm, D., \& Aharonov, Y. (1957). Discussion of experimental proof
for the paradox of Einstein, Rosen and Podolski. \emph{Physical Review,
108}, 1070-1076.

Bohr, N. (1935). Can quantum-mechanical description of physical reality
be considered complete? \emph{Physical Review, 48}, 696-702.

Budroni, C. (2016). \emph{Temporal Quantum Correlations and Hidden
Variable Models.} Springer: Heidelberg.

Budroni, C. \& Emary, C. (2014). Temporal quantum correlations and
Leggett-Garg inequalities in multilevel systems. \emph{Physical Review
Letters, 113}, 050401.

Bruza, P. D., Kitto, K., Nelson, D., \& McEvoy, C. (2009). Is there
something quantum-like about the human mental lexicon? \emph{Journal
of Mathematical Psychology, 53}, 362\textendash 377.

Bruza, P. D., Kitto, K., Ramm, B. J., \& Sitbon, L. (2015). A probabilistic
framework for analysing the compositionality of conceptual combinations.
\emph{Journal of Mathematical Psychology, 67}, 26-38.

Cabello, A. (2013). Simple explanation of the quantum violation of
a fundamental inequality. \emph{Physical Review Letters, 110}, 060402.

Cepeda Cuervo, E., Aguilar, W., Cervantes, V. H., Corrales, M., Díaz,
I. \& Rodríguez, D. (2008). Intervalos de confianza e intervalos de
credibilidad para una proporción. \emph{Revista Colombiana de Estadística,
31, }211-228.

Cereceda, J. (2000). Quantum mechanical probabilities and general
probabilistic constraints for Einstein\textendash Podolsky\textendash Rosen\textendash Bohm
experiments. \emph{Foundations of Physics Letters, 13}, 427\textendash 442.

Cervantes, V. H., \& Dzhafarov, E. N. (2017a). Exploration of contextuality
in a psychophysical double-detection experiment. \emph{Lecture Notes
in Computer Science, 10106}, 182-193.

Cervantes, V. H., \& Dzhafarov, E. N. (2017b). Advanced analysis of
quantum contextuality in a psychophysical double-detection experiment.
\emph{Journal of Mathematical Psychology, 79}, 77-84.

Cervantes, V. H., \& Dzhafarov, E. N. (2018). Snow Queen is evil and
beautiful: Experimental evidence for probabilistic contextuality in
human choices. \emph{Decision, 5}, 193-204.

Clauser, J. F., \& Horne, M. A. (1974). Experimental consequences
of objective local theories. \emph{Physical Review D, 10}, 526-535.

Clauser, J. F. , Horne, M. A. , Shimony, A., \& Holt, R. A. (1969).
Proposed experiment to test local hidden-variable theories. \emph{Physical
Review Letters, 23}, 880-884.

Clopper, C. \& Pearson, E. S. (1934). The use of confidence or fiducial
limits illustrated in the case of the binomial. \emph{Biometrika,
26}, 404\textendash 413.

Dzhafarov, E. N. (2003). Selective influence through conditional independence.
\emph{Psychometrika, 68}, 7-26.

Dzhafarov, E. N. (2017). Replacing nothing with something special:
Contextuality-by-Default and dummy measurements. In A. Khrennikov
\& T. Bourama (Eds.). \emph{Quantum Foundations, Probability and Information
}(pp. 39-44). Berlin: Springer.

Dzhafarov, E. N., Cervantes, V. H., \& Kujala, J. V. (2017). Contextuality
in canonical systems of random variables. \emph{Philosophical Transactions
of the Royal Society A, 375}, 20160389.

Dzhafarov, E. N., \& Kon, M. (2018). On universality of classical
probability with contextually labeled random variables. \emph{Journal
of Mathematical Psychology, 85}, 17-24

Dzhafarov, E. N., \& Kujala, J. V. (2014). Selective influences, marginal
selectivity, and Bell/CHSH inequalities. \emph{Topics in Cognitive
Science, 6, }121\textendash 128.

Dzhafarov, E. N., \& Kujala, J. V. (2016a). Context-content systems
of random variables: The contextuality-by-default theory. \emph{Journal
of Mathematical Psychology, 74}, 11-33.

Dzhafarov, E. N., \& Kujala, J. V. (2016b). Probability, random variables,
and selectivity. In W. Batchelder, H. Colonius, E. N. Dzhafarov, \&
J. Myung (Eds). \emph{The New Handbook of Mathematical Psychology
}(pp. 85-150). Cambridge University Press.

Dzhafarov, E. N. \& Kujala, J. V. (2017a). Probabilistic foundations
of contextuality. \emph{Fortschritte der Physik, 65}, 1-11.

Dzhafarov, E. N. \& Kujala, J. V. (2017b). Contextuality-by-Default
2.0: Systems with binary random variables. \emph{Lecture Notes Computer
Sciences, 10106}, 16-32.

Dzhafarov, E. N., \& Kujala, J. V. (2018). Contextuality analysis
of the double slit experiment (with a glimpse into three slits). \emph{Entropy,
20}, 278.

Dzhafarov, E. N., Kujala, J. V., Cervantes, V. H., Zhang, R., \& Jones,
M. (2016). On contextuality in behavioral data. \emph{Philosophical
Transactions of the Royal Society A, 374}, 20150234.

Dzhafarov, E. N., Kujala, J. V., \& Larsson, J.-Å. (2015). Contextuality
in three types of quantum-mechanical systems. \emph{Foundations of
Physics, 7,} 762-782.

Dzhafarov, E. N., Zhang, R., \& Kujala, J. V. (2015). Is there contextuality
in behavioral and social systems? \emph{Philosophical Transactions
of the Royal Society A, 374}, 20150099.

Einstein, A., Podolsky, B., \& Rosen, N. (1935). Can quantum-mechanical
description of physical reality be considered complete? \emph{Physical
Review, 47}, 777-780.

Fine, A. (1982a). Hidden variables, joint probability, and the Bell
inequalities. \emph{Physical Review Letters, 48}, 291-295.

Fine, A. (1982b). Joint distributions, quantum correlations, and commuting
observables. \emph{Journal of Mathematical Physics, 23},1306-1310.

Filk, T. (2015). It is the theory which decides what we can observe.
In E. N. Dzhafarov, S. Jordan, R. Zhang, \& V. Cervantes (Eds). Cont\emph{extuality
from Quantum Physics to Psychology }(pp. 77-92). New Jersey: World
Scientific.

Filk, T. (2016). A mechanical model of a PR-Box. arXiv:1507.06789.

Griffiths, R. B. (2017). What quantum measurements measure. \emph{Physical
Review A, 96}, 032110.

Klyachko, A. A., Can, M. A., Binicioglu, S., \& Shumovsky, A. S. (2008).
A simple test for hidden variables in spin-1 system. \emph{Physical
Review Letters, 101}, 020403.
\end{onehalfspace}

Khrennikov, A. (2000a). Non-Kolmogorov probability models and modified
Bell's inequality. \emph{Journal of Mathematical Physics, 41}, 1768-1777.

Khrennikov, A. (2000b). A perturbation of CHSH inequality induced
by fluctuations of ensemble distributions. \emph{Journal of Mathematical
Physics, 41}, 5934-5944.

Khrennikov, A. (2001). Contextualist viewpoint to Greenberger-Horne-Zeilinger
paradox. \emph{Physical Letters A, 278}, 307-314.

Khrennikov, A. (2008). Bell-Boole inequality: Nonlocality or probabilistic
incompatibility of random variables? \emph{Entropy, 10, }19-32.

\begin{onehalfspace}
Kochen, S., \& Specker, E. P. (1967). The problem of hidden variables
in quantum mechanics. \emph{Journal of Mathematics and Mechanics,
17}, 59\textendash 87.

Kujala, J. V., \& Dzhafarov, E. N. (2015). Probabilistic contextuality
in EPR/Bohm-type systems with signaling allowed. In E. N. Dzhafarov,
S. Jordan, R. Zhang, \& V. Cervantes (Eds). \emph{Contextuality from
Quantum Physics to Psychology }(pp. 287-308). New Jersey: World Scientific.

Kujala, J. V., \& Dzhafarov, E. N. (2016). Proof of a conjecture on
contextuality in cyclic systems with binary variables. \emph{Foundations
of Physics, 46}, 282-299.

Kurzynski, P., Ramanathan, R., \& Kaszlikowski, D. (2012). Entropic
test of quantum contextuality. \emph{Physical Review Letters, 109},
020404.

Lapkiewicz, R., Li, P., Schaeff, C., Langford, N. K., Ramelow, S.,
Wie\'{s}niak, M., \& Zeilinger, A. (2011). Experimental non-classicality
of an indivisible quantum system. \emph{Nature, 474}, 490\textendash 493.

Leggett, A., \& Garg, A. (1985). Quantum mechanics versus macroscopic
realism: Is the flux there when nobody looks? \emph{Physical Review
Letters, 54}, 857.

Liang, Y.-C., Spekkens, R. W., \& Wiseman, H. M. (2011). Specker\textquoteright s
parable of the overprotective seer: A road to contextuality, nonlocality
and complementarity. \emph{Physics Reports, 506}, 1-39.

Ramanathan, R., Soeda, A., Kurzynski, P., \& Kaszlikowski, D. (2012).
Generalized monogamy of contextual inequalities from the no-disturbance
principle. \emph{Physical Review Letters, 109}, 050404.

Suppes, P., \& Zanotti, M. (1981). When are probabilistic explanations
possible? \emph{Synthese, 48}, 191-199.

Thorisson, H. (2000). \emph{Coupling, Stationarity, and Regeneration}.
New York: Springer.

Wang, Z., \& Busemeyer, J. R. (2013). A quantum question order model
supported by empirical tests of an a priori and precise prediction.
\emph{Topics in Cognitive Science, 5}, 689\textendash 710.

Wang, Z., Solloway, T., Shiffrin, R. M., \& Busemeyer, J. R. (2014).
Context effects produced by question orders reveal quantum nature
of human judgments. \emph{Proceedings of the National Academy of Sciences,
111}, 9431-9436.

Zhang, R., \& Dzhafarov, E. N. (2016). Testing contextuality in cyclic
psychophysical systems of high ranks. \emph{Lecture Notes in Computer
Science, 10106}, 151-162.
\end{onehalfspace}

\newpage{}

\section*{Appendix: Data tables}

Table A1: Empirical estimates of marginal distributions for the conteNt-conteXt
system in Fig. \ref{fig:positionSystem} for participant P1.

$\begin{tabular}{lrrrrrrcrrrrr}
\hline  P1  &   & \multicolumn{5}{c}{Left response } &   & \multicolumn{5}{c}{Right response }\\
 \cline{3-7} \cline{9-13} \\
 Context  &  Trials  &  Center  &  Left  &  Right  &  Up  &  Down  &   &  Center  &  Left  &  Right  &  Up  &  Down\\
\hline  \ensuremath{cc}  &  336  &  .318  &  .521  &  .000  &  .155  &  .006  &   &  .235  &  .455  &  .000  &  .310  &  .000\\
 \ensuremath{cl}  &  334  &  .213  &  .656  &  .000  &  .132  &  .000  &   &  .015  &  .871  &  .000  &  .108  &  .006\\
 \ensuremath{cr}  &  336  &  .390  &  .435  &  .000  &  .155  &  .021  &   &  .601  &  .060  &  .018  &  .318  &  .003\\
 \ensuremath{cu}  &  336  &  .298  &  .554  &  .000  &  .143  &  .006  &   &  .021  &  .149  &  .000  &  .830  &  .000\\
 \ensuremath{cd}  &  336  &  .265  &  .574  &  .000  &  .152  &  .009  &   &  .271  &  .613  &  .000  &  .030  &  .086\\
 \ensuremath{lc}  &  334  &  .036  &  .931  &  .000  &  .027  &  .006  &   &  .195  &  .527  &  .000  &  .278  &  .000\\
 \ensuremath{ll}  &  335  &  .024  &  .928  &  .000  &  .042  &  .006  &   &  .021  &  .860  &  .000  &  .119  &  .000\\
 \ensuremath{lr}  &  335  &  .051  &  .913  &  .003  &  .030  &  .003  &   &  .558  &  .122  &  .018  &  .299  &  .003\\
 \ensuremath{lu}  &  335  &  .054  &  .904  &  .000  &  .033  &  .009  &   &  .042  &  .176  &  .003  &  .779  &  .000\\
 \ensuremath{ld}  &  334  &  .042  &  .910  &  .003  &  .042  &  .003  &   &  .314  &  .605  &  .000  &  .024  &  .057\\
 \ensuremath{rc}  &  333  &  .763  &  .081  &  .033  &  .117  &  .006  &   &  .246  &  .483  &  .000  &  .270  &  .000\\
 \ensuremath{rl}  &  334  &  .605  &  .159  &  .051  &  .183  &  .003  &   &  .018  &  .859  &  .000  &  .120  &  .003\\
 \ensuremath{rr}  &  335  &  .782  &  .048  &  .024  &  .137  &  .009  &   &  .591  &  .042  &  .021  &  .346  &  .000\\
 \ensuremath{ru}  &  336  &  .685  &  .077  &  .027  &  .202  &  .009  &   &  .045  &  .083  &  .000  &  .872  &  .000\\
 \ensuremath{rd}  &  335  &  .701  &  .075  &  .036  &  .179  &  .009  &   &  .322  &  .555  &  .000  &  .033  &  .090\\
 \ensuremath{uc}  &  335  &  .116  &  .269  &  .003  &  .612  &  .000  &   &  .200  &  .457  &  .000  &  .343  &  .000\\
 \ensuremath{ul}  &  336  &  .062  &  .345  &  .000  &  .592  &  .000  &   &  .021  &  .872  &  .003  &  .101  &  .003\\
 \ensuremath{ur}  &  334  &  .156  &  .216  &  .000  &  .629  &  .000  &   &  .581  &  .051  &  .027  &  .335  &  .006\\
 \ensuremath{uu}  &  334  &  .084  &  .260  &  .000  &  .656  &  .000  &   &  .033  &  .108  &  .000  &  .859  &  .000\\
 \ensuremath{ud}  &  335  &  .096  &  .191  &  .000  &  .713  &  .000  &   &  .343  &  .558  &  .000  &  .033  &  .066\\
 \ensuremath{dc}  &  335  &  .337  &  .478  &  .000  &  .006  &  .179  &   &  .242  &  .460  &  .000  &  .296  &  .003\\
 \ensuremath{dl}  &  334  &  .237  &  .599  &  .000  &  .006  &  .159  &   &  .012  &  .880  &  .000  &  .108  &  .000\\
 \ensuremath{dr}  &  336  &  .312  &  .449  &  .000  &  .009  &  .229  &   &  .589  &  .054  &  .027  &  .330  &  .000\\
 \ensuremath{du}  &  335  &  .310  &  .504  &  .000  &  .015  &  .170  &   &  .030  &  .116  &  .000  &  .854  &  .000\\
 \ensuremath{dd}  &  335  &  .346  &  .451  &  .000  &  .006  &  .197  &   &  .370  &  .549  &  .000  &  .012  &  .069\\
\hline  
\end{tabular}$

\newpage Table A2. Empirical estimates of marginal distributions for
the conteNt-conteXt system in Fig. \ref{fig:positionSystem} for participant
P2.

$\begin{tabular}{lrrrrrrcrrrrr}
\hline  P2  &   & \multicolumn{5}{c}{Left response } &   & \multicolumn{5}{c}{Right response }\\
 \cline{3-7} \cline{9-13} \\
 Context  &  Trials  &  Center  &  Left  &  Right  &  Up  &  Down  &   &  Center  &  Left  &  Right  &  Up  &  Down\\
\hline  \ensuremath{cc}  &  336  &  .616  &  .062  &  .039  &  .226  &  .057  &   &  .560  &  .062  &  .164  &  .202  &  .012\\
 \ensuremath{cl}  &  336  &  .586  &  .080  &  .033  &  .268  &  .033  &   &  .265  &  .604  &  .015  &  .107  &  .009\\
 \ensuremath{cr}  &  336  &  .586  &  .045  &  .062  &  .259  &  .048  &   &  .185  &  .000  &  .720  &  .071  &  .024\\
 \ensuremath{cu}  &  336  &  .607  &  .083  &  .033  &  .220  &  .057  &   &  .131  &  .062  &  .089  &  .717  &  .000\\
 \ensuremath{cd}  &  336  &  .580  &  .054  &  .024  &  .304  &  .039  &   &  .348  &  .033  &  .086  &  .024  &  .509\\
 \ensuremath{lc}  &  336  &  .223  &  .604  &  .000  &  .134  &  .039  &   &  .610  &  .092  &  .119  &  .152  &  .027\\
 \ensuremath{ll}  &  336  &  .214  &  .583  &  .003  &  .164  &  .036  &   &  .274  &  .548  &  .021  &  .134  &  .024\\
 \ensuremath{lr}  &  336  &  .223  &  .586  &  .009  &  .158  &  .024  &   &  .220  &  .006  &  .682  &  .065  &  .027\\
 \ensuremath{lu}  &  336  &  .310  &  .527  &  .000  &  .128  &  .036  &   &  .149  &  .042  &  .089  &  .720  &  .000\\
 \ensuremath{ld}  &  336  &  .226  &  .557  &  .003  &  .179  &  .036  &   &  .333  &  .039  &  .098  &  .021  &  .509\\
 \ensuremath{rc}  &  336  &  .339  &  .003  &  .443  &  .176  &  .039  &   &  .548  &  .086  &  .158  &  .173  &  .036\\
 \ensuremath{rl}  &  336  &  .318  &  .012  &  .432  &  .205  &  .033  &   &  .247  &  .631  &  .018  &  .080  &  .024\\
 \ensuremath{rr}  &  336  &  .310  &  .003  &  .429  &  .229  &  .030  &   &  .140  &  .018  &  .696  &  .116  &  .030\\
 \ensuremath{ru}  &  336  &  .336  &  .000  &  .467  &  .158  &  .039  &   &  .170  &  .033  &  .074  &  .720  &  .003\\
 \ensuremath{rd}  &  336  &  .351  &  .000  &  .405  &  .211  &  .033  &   &  .381  &  .054  &  .095  &  .030  &  .440\\
 \ensuremath{uc}  &  336  &  .146  &  .018  &  .015  &  .818  &  .003  &   &  .646  &  .048  &  .134  &  .137  &  .036\\
 \ensuremath{ul}  &  336  &  .131  &  .030  &  .015  &  .821  &  .003  &   &  .345  &  .545  &  .012  &  .068  &  .030\\
 \ensuremath{ur}  &  336  &  .146  &  .030  &  .006  &  .815  &  .003  &   &  .235  &  .000  &  .688  &  .057  &  .021\\
 \ensuremath{uu}  &  336  &  .167  &  .021  &  .018  &  .795  &  .000  &   &  .196  &  .036  &  .128  &  .637  &  .003\\
 \ensuremath{ud}  &  336  &  .137  &  .015  &  .009  &  .836  &  .003  &   &  .390  &  .024  &  .068  &  .015  &  .503\\
 \ensuremath{dc}  &  336  &  .354  &  .030  &  .036  &  .021  &  .560  &   &  .539  &  .057  &  .143  &  .229  &  .033\\
 \ensuremath{dl}  &  336  &  .366  &  .039  &  .024  &  .018  &  .554  &   &  .220  &  .583  &  .006  &  .158  &  .033\\
 \ensuremath{dr}  &  336  &  .375  &  .039  &  .006  &  .009  &  .571  &   &  .199  &  .003  &  .661  &  .119  &  .018\\
 \ensuremath{du}  &  336  &  .393  &  .018  &  .033  &  .021  &  .536  &   &  .122  &  .027  &  .065  &  .786  &  .000\\
 \ensuremath{dd}  &  336  &  .360  &  .039  &  .024  &  .036  &  .542  &   &  .393  &  .048  &  .116  &  .021  &  .423\\
\hline  
\end{tabular}$

\newpage Table A3. Empirical estimates of marginal distributions for
the conteNt-conteXt system in Fig. \ref{fig:positionSystem} for participant
P3.

$\begin{tabular}{lrrrrrrcrrrrr}
\hline  P3  &   & \multicolumn{5}{c}{Left response } &   & \multicolumn{5}{c}{Right response }\\
 \cline{3-7} \cline{9-13} \\
 Context  &  Trials  &  Center  &  Left  &  Right  &  Up  &  Down  &   &  Center  &  Left  &  Right  &  Up  &  Down\\
\hline  \ensuremath{cc}  &  336  &  .738  &  .092  &  .012  &  .143  &  .015  &   &  .634  &  .149  &  .015  &  .188  &  .015\\
 \ensuremath{cl}  &  337  &  .801  &  .053  &  .030  &  .110  &  .006  &   &  .027  &  .955  &  .000  &  .018  &  .000\\
 \ensuremath{cr}  &  336  &  .762  &  .098  &  .021  &  .107  &  .012  &   &  .321  &  .003  &  .631  &  .036  &  .009\\
 \ensuremath{cu}  &  336  &  .768  &  .086  &  .033  &  .098  &  .015  &   &  .128  &  .060  &  .000  &  .812  &  .000\\
 \ensuremath{cd}  &  335  &  .785  &  .081  &  .021  &  .101  &  .012  &   &  .648  &  .081  &  .009  &  .003  &  .260\\
 \ensuremath{lc}  &  337  &  .056  &  .935  &  .000  &  .009  &  .000  &   &  .700  &  .104  &  .039  &  .151  &  .006\\
 \ensuremath{ll}  &  336  &  .060  &  .929  &  .000  &  .012  &  .000  &   &  .045  &  .929  &  .000  &  .027  &  .000\\
 \ensuremath{lr}  &  337  &  .053  &  .929  &  .000  &  .015  &  .003  &   &  .288  &  .000  &  .680  &  .033  &  .000\\
 \ensuremath{lu}  &  337  &  .059  &  .917  &  .000  &  .021  &  .003  &   &  .148  &  .059  &  .006  &  .786  &  .000\\
 \ensuremath{ld}  &  336  &  .051  &  .938  &  .000  &  .012  &  .000  &   &  .676  &  .054  &  .015  &  .006  &  .250\\
 \ensuremath{rc}  &  336  &  .336  &  .000  &  .649  &  .012  &  .003  &   &  .658  &  .125  &  .024  &  .185  &  .009\\
 \ensuremath{rl}  &  337  &  .335  &  .009  &  .635  &  .021  &  .000  &   &  .027  &  .935  &  .000  &  .039  &  .000\\
 \ensuremath{rr}  &  336  &  .312  &  .000  &  .670  &  .012  &  .006  &   &  .298  &  .000  &  .667  &  .036  &  .000\\
 \ensuremath{ru}  &  337  &  .332  &  .000  &  .653  &  .015  &  .000  &   &  .142  &  .071  &  .000  &  .783  &  .003\\
 \ensuremath{rd}  &  336  &  .280  &  .000  &  .699  &  .015  &  .006  &   &  .658  &  .074  &  .021  &  .012  &  .235\\
 \ensuremath{uc}  &  336  &  .164  &  .033  &  .003  &  .801  &  .000  &   &  .699  &  .134  &  .021  &  .137  &  .009\\
 \ensuremath{ul}  &  336  &  .143  &  .065  &  .003  &  .789  &  .000  &   &  .042  &  .943  &  .000  &  .012  &  .003\\
 \ensuremath{ur}  &  336  &  .134  &  .033  &  .003  &  .830  &  .000  &   &  .327  &  .000  &  .631  &  .033  &  .009\\
 \ensuremath{uu}  &  336  &  .202  &  .033  &  .003  &  .762  &  .000  &   &  .164  &  .062  &  .000  &  .774  &  .000\\
 \ensuremath{ud}  &  337  &  .172  &  .021  &  .003  &  .804  &  .000  &   &  .668  &  .080  &  .015  &  .000  &  .237\\
 \ensuremath{dc}  &  335  &  .603  &  .021  &  .012  &  .000  &  .364  &   &  .618  &  .137  &  .009  &  .230  &  .006\\
 \ensuremath{dl}  &  337  &  .626  &  .030  &  .027  &  .000  &  .318  &   &  .030  &  .950  &  .000  &  .021  &  .000\\
 \ensuremath{dr}  &  337  &  .644  &  .030  &  .021  &  .000  &  .306  &   &  .329  &  .000  &  .635  &  .033  &  .003\\
 \ensuremath{du}  &  337  &  .638  &  .030  &  .015  &  .000  &  .318  &   &  .151  &  .080  &  .003  &  .766  &  .000\\
 \ensuremath{dd}  &  336  &  .619  &  .039  &  .012  &  .003  &  .327  &   &  .708  &  .068  &  .009  &  .006  &  .208\\
\hline  
\end{tabular}$
\end{document}